\documentclass[11pt]{elsart}
\usepackage[dvips]{graphicx}
\usepackage{amsmath}
\usepackage{amssymb}

\begin{document}
\begin{frontmatter}

\title{A unified model for Sierpinski networks with scale-free scaling and small-world effect}

\author[label1]{Jihong Guan}
\ead{jhguan@mail.tongji.edu.cn}
\author[label1]{Yuewen Wu}
\author[lable2,label3]{Zhongzhi Zhang}
\ead{zhangzz@fudan.edu.cn}
\author[lable2,label3]{Shuigeng Zhou}
\ead{sgzhou@fudan.edu.cn}
\author[lable2,label3]{Yonghui Wu}
\ead{yhwu@fudan.edu.cn}

\address[label1]{Department of Computer Science and Technology, Tongji University, 4800 Cao'an Road, Shanghai 201804, China}
\address[lable2]{School of Computer Science, Fudan
University, Shanghai 200433, China}
\address[label3]{Shanghai Key Lab of Intelligent Information
Processing, Fudan University, Shanghai 200433, China}

\begin{abstract}
In this paper, we propose an evolving Sierpinski gasket, based on
which we establish a model of evolutionary Sierpinski networks
(ESNs) that unifies deterministic Sierpinski network [Eur. Phys. J.
B {\bf 60}, 259 (2007)] and random Sierpinski network [Eur. Phys. J.
B {\bf 65}, 141 (2008)] to the same framework. We suggest an
iterative algorithm generating the ESNs. On the basis of the
algorithm, some relevant properties of presented networks are
calculated or predicted analytically. Analytical solution shows that
the networks under consideration follow a power-law degree
distribution, with the distribution exponent continuously tuned in a
wide range. The obtained accurate expression of clustering
coefficient, together with the prediction of average path length
reveals that the ESNs possess small-world effect. All our
theoretical results are successfully contrasted by numerical
simulations. Moreover, the evolutionary prisoner's dilemma game is
also studied on some limitations of the ESNs, i.e., deterministic
Sierpinski network and random Sierpinski network.
\\


\end{abstract}


\date{\today}
\end{frontmatter}


\section{Introduction}

In the last few years, complex networks have attracted a growing
interest from a wide circle of
researchers~\cite{AlBa02,DoMe02,Ne03,BoLaMoChHw06}. The reason for
this boom is that complex networks describe various systems in
nature and society, such as the World Wide Web (WWW), the Internet,
collaboration networks, and sexual network, and so on. Extensive
empirical studies have revealed that real-life systems have in
common at least two striking statistical properties: power-law
degree distribution~\cite{BaAl99}, small-world effect~\cite{WaSt98}
including small average path length (APL) and high clustering
coefficient. In order to mimic real-word systems with above
mentioned common characteristics, a wide variety of models have been
proposed~\cite{AlBa02,DoMe02,Ne03,BoLaMoChHw06}. At present, it is
still an active direction to construct models reproducing the
structure and statistical characteristics of real systems.

In our previous papers, on the basis of the well-known Sierpinski
fractal (or Sierpinski gasket), we have proposed a deterministic
network called deterministic Sierpinski network
(DSN)~\cite{ZhZhChGu07}, and a stochastic network named random
Sierpinski network (RSN)~\cite{ZhZhSuZoGu08}, respectively. Both the
DSN and RSN possess good topological properties observed in some
real systems. In this paper, we suggest a general scenario for
constructing evolutionary Sierpinski networks (ESNs) controlled by a
parameter $q$. The ESNs can also result from Sierpinski gasket and
unify the DSN and RSN to the same framework, i.e., the DSN and RSN
are special cases of RSNs. The ESNs have a power-law degree
distribution, a very large clustering coefficient, and a small
intervertex separation. The degree exponent of ESNs is changeable
between $2$ and $3$. Moreover, we introduce a generating algorithm
for the ESNs which can realize the construction of our networks. In
the end, the cooperation behavior of the evolutionary prisoner's
dilemma game on two limiting cases (i.e., DSN and RSN) of the ESNs
is discussed.

\begin{figure}
\begin{center}
\includegraphics[width=0.6\textwidth]{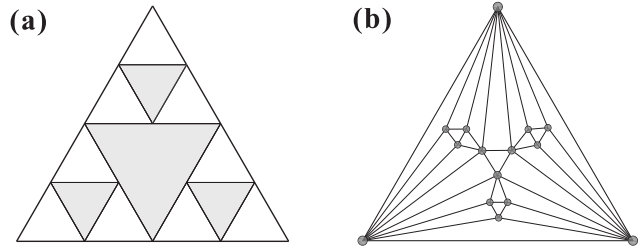}
\end{center}
\caption[kurzform]{\label{network} The first two stages of
construction of the Sierpinski gasket (a) and its corresponding
network (b).}
\end{figure}

\section{Brief introduction to deterministic and random Sierpinski networks}

We first introduce Sierpinski gasket, which is also known as
Sierpinski triangle. The classical Sierpinski gasket denoted as
$S_{t}$ after $t$ generations, is constructed as
follows~\cite{Si15,Re94}: start with an equilateral triangle, and
denote this initial configuration as $S_{0}$. Perform a bisection of
the sides forming four small copies of the original triangle, and
remove the interior triangles to get $S_{1}$. Repeat this procedure
recursively in the three remaining copies to obtain $S_{2}$, see
Fig.~\ref{network}(a). In the infinite $t$ limit, we obtain the
famous Sierpinski gasket $S_{t}$. From Sierpinski gasket we can
easily construct a network, called deterministic Sierpinski network,
with sides of the removed triangles mapped to nodes and contact to
edges between nodes~\cite{ZhZhChGu07}. For uniformity, the three
sides of the initial equilateral triangle at step 0 also correspond
to three different nodes. Figure~\ref{network}(b) shows a network
based on $S_{2}$.

\begin{figure}
\begin{center}
\includegraphics[width=0.4\textwidth]{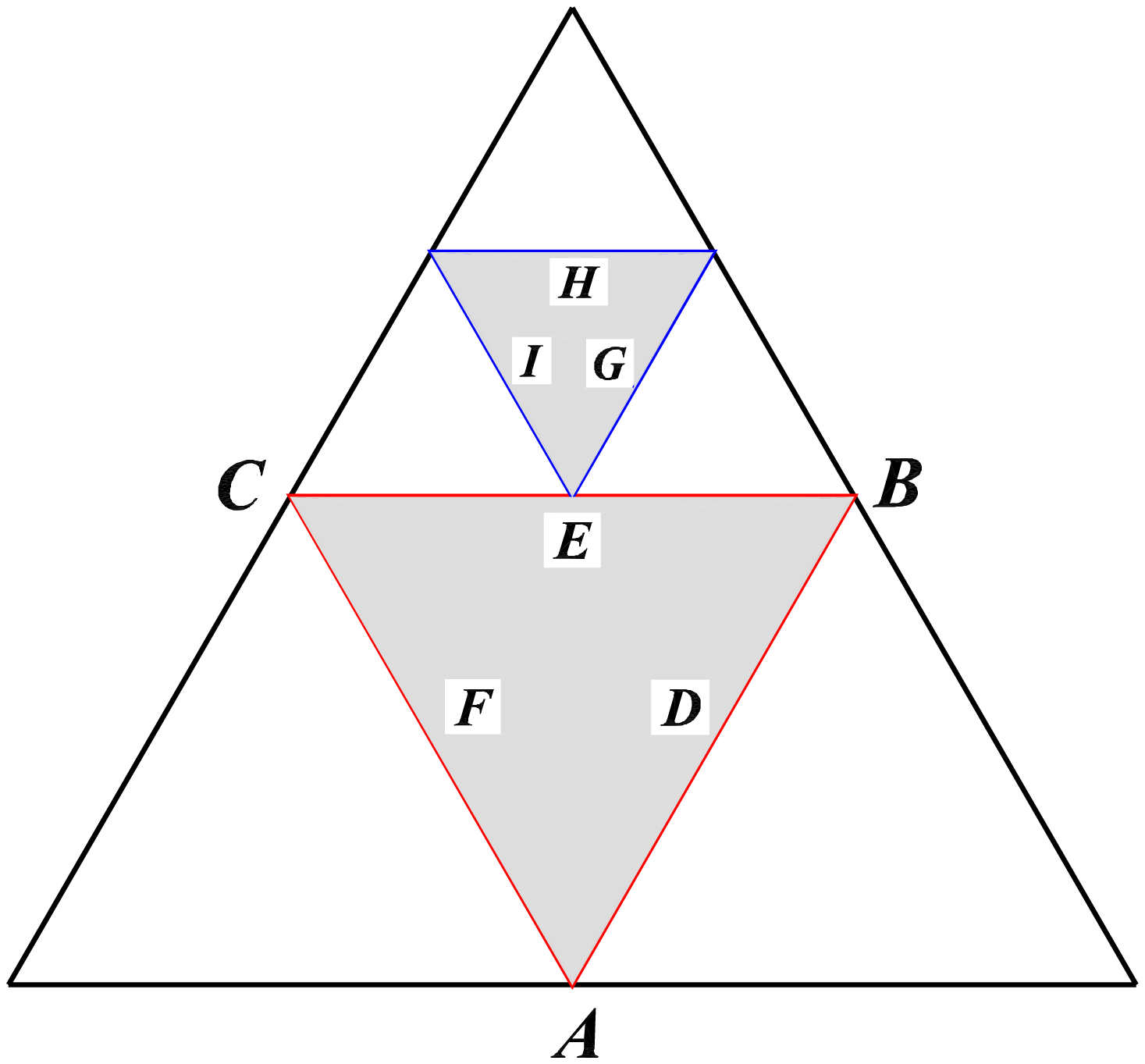}
\includegraphics[width=0.36\textwidth]{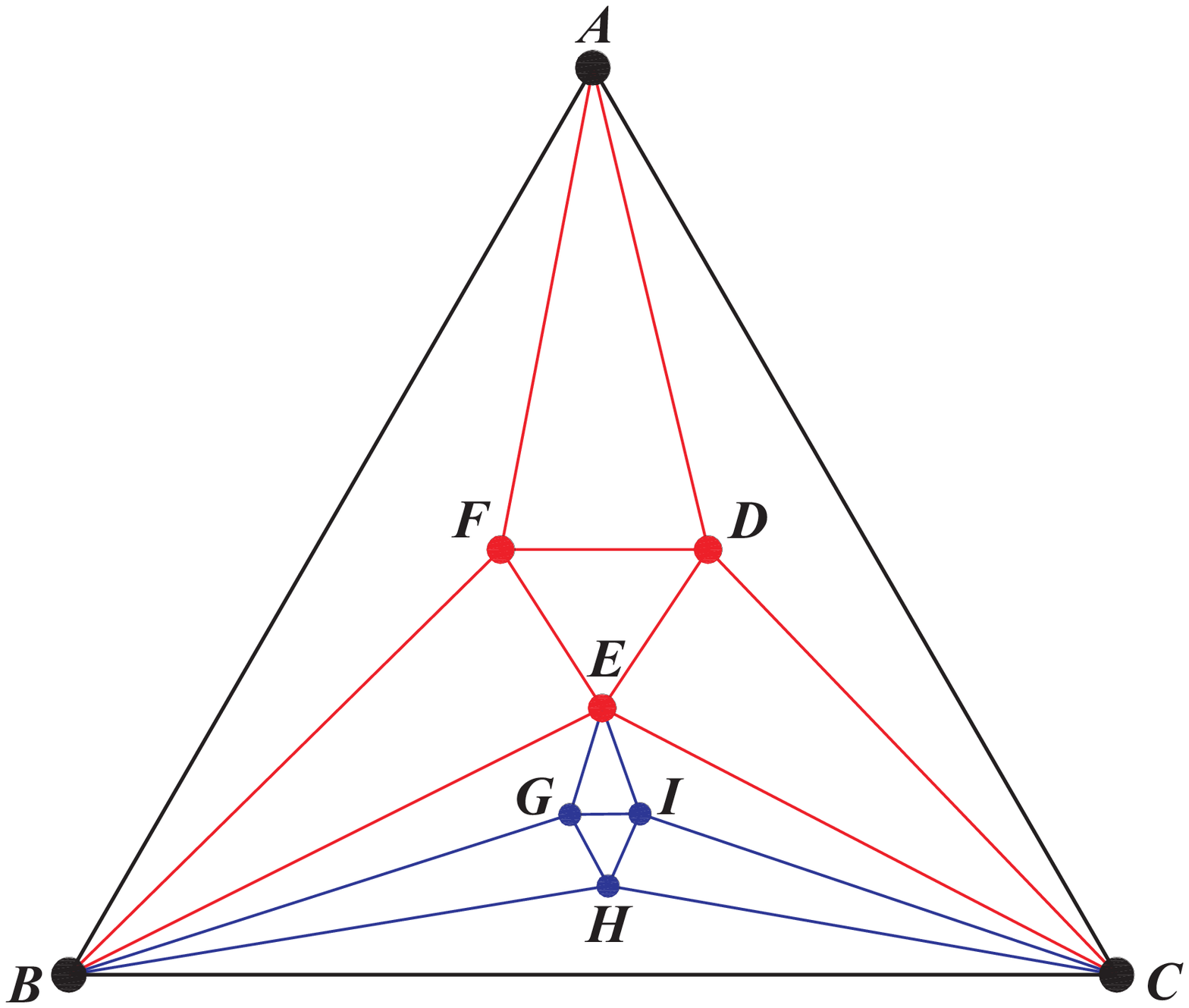}
\end{center}
\caption[kurzform]{\label{web} (Color online) The sketch maps for
the construction of random Sierpinski gasket (left panel) and its
corresponding network (right panel).}
\end{figure}

Analogously, one can construct the random Sierpinski
network~\cite{ZhZhSuZoGu08} derived from the stochastic Sierpinski
gasket, which is a random variant of the deterministic Sierpinski
gasket. The initial configuration of the random  Sierpinski gasket
is the same as the deterministic Sierpinski triangle. Then in each
of the subsequent generations, an equilateral triangle is chosen
randomly, for which bisection and removal are performed to form
three small copies of it. The sketch map for the random fractal is
shown in the left panel of Fig.~\ref{web}. From this fractal we can
easily establish the random Sierpinski network with sides of the
removed triangles mapped to nodes and contact to links between
nodes. The right panel of Fig.~\ref{web} gives a network derived
from the random Sierpinski gasket.

\section{Unifying model and its iterative algorithm}

In this section, we introduce an evolving unified model for the
deterministic and random Sierpinski networks. First we give a new
variation, called evolving Sierpinski gasket (ESG), for the
Sierpinski gasket. The initial configuration of the ESG is the same
as the deterministic Sierpinski gasket. Then in each of the
subsequent generations, for each equilateral triangle, with
probability $q$, bisection and removal are performed to form three
small copies of it. In the infinite generation limit, the ESG is
obtained. In a special case $q=1$, the ESG is reduced to the classic
deterministic Sierpinski gasket. If $q$ approaches but is not equal
to $0$, it coincides with the random Sierpinski gasket described in
Ref.~\cite{ZhZhSuZoGu08}. The proposed unified model is derived from
this ESG: nodes represent the sides of the removed triangles and
edges correspond to contact relationship. As in the construction of
the deterministic and random Sierpinski
networks~\cite{ZhZhChGu07,ZhZhSuZoGu08}, the three sides of the
initial equilateral triangle (at step 0) of the ESG are also mapped
to three different nodes.

In the construction process of the ESG, for each equilateral
triangle at arbitrary generation, once we perform a bisection of its
sides and remove the central down pointing triangle, three copies of
it are formed. When building the unifying network model, it is
equivalent that for each group of three newly-added nodes, three new
triangles are generated, which may create new nodes in the
subsequent generations. According to this, we can introduce an
iterative algorithm to create the ESNs. Using the proposed algorithm
one can write a computer program conveniently to simulate the
networks and study their properties.

\begin{figure}
\begin{center}
\includegraphics[width=0.50\textwidth]{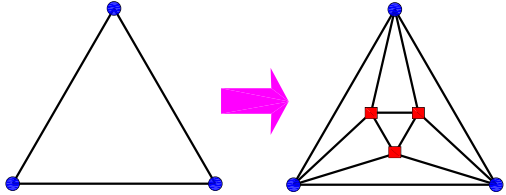}
\end{center}
\caption[kurzform]{\label{iterative} (Color online) Iterative
construction method for the network. }
\end{figure}

We denote the ESNs after $t$ ($t\geq0$) iterations by $EW(t)$, then
the proposed algorithm to create ESNs is as follows. Initially
$(t=0)$, $EW(0)$ has three nodes forming a triangle. At step 1, with
probability $q$, we add three nodes into the original triangle.
These three new nodes are connected to one another shaping a new
triangle, and both ends of each edge of the new triangle are linked
to a node of the original triangle. Thus we obtain $EW(1)$, see
Fig.~\ref{iterative}. For $t\geq 1$, $EW(t)$ is obtained from
$EW(t-1)$. For the convenience of description, we give the following
definition: for each existing  triangle in $EW(t-1)$, if there is no
node in its interior and among its three nodes there is only one
youngest node (i.e., the other two are strictly elder than it), we
call it an active triangle (with the initial triangle as an
exception). At step $t-1$, for each existing active triangle, with
probability $q$ it is replaced by the connected cluster on the right
of Fig.~\ref{iterative}, then $EW(t)$ is produced. The growing
process is repeated until the network reaches a desired order. When
$q=1$, the network is exactly the same as the DSN~\cite{ZhZhChGu07}.
If $q<1$, the network grows randomly. Especially, as $q$ approaches
zero and does not equal zero, the network is reduced to the RSN
studied in detail in Ref.~\cite{ZhZhSuZoGu08}.

Next we compute the order (number of all nodes) and size (number of
all edges) of $EW(t)$. Denote $L_\Delta(t)$ as the number of active
triangles at step $t$. Then, $L_\Delta(0)=1$. By construction, we
can easily derive that $L_\Delta(t)=(1+2q)^{t}$. Let $L_v(t)$ and
$L_e(t)$ be the number of nodes and edges created at step $t$,
respectively. Note that each active triangle in $EW(t-1)$ will (see
Fig.~\ref{iterative}) lead to three new nodes and nine new edges in
$EW(t)$ with probability $q$. Then, at step $1$, we add expected
$L_{v}(1)=3q$ new nodes and $L_{e}(1)=9q$ new edges to $EW(0)$.
After simple calculations, one can obtain that at step $t_{i}$
$(t_{i}>1)$ the number of newly-born nodes and edges is
$L_{v}(t_{i})=3q\,L_\Delta(t_i-1)=3q(1+2q)^{t_{i}-1}$ and
$L_{e}(t_{i})=9q\,L_\Delta(t_i-1)=9q(1+2q)^{t_{i}-1}$, respectively.
Thus the average number of total nodes $N_{t}$ and edges $E_{t}$
present at step $t$ is
\begin{equation}\label{Nt}
N_{t} =3+\sum_{t_{i}=1}^{t}L_{v}(t_{i})=\frac{3(1+2q)^{t}+3}{2}
\end{equation}
and
\begin{equation}\label{Et}
E_{t}=3+\sum_{t_{i}=1}^{t}L_{e}(t_{i})=\frac{9(1+2q)^{t}-3}{2},
\end{equation}
respectively. So for large $t$, the average degree
$\overline{k_{t}}=\frac{2E_{t}}{N_{t}}$ is approximately $6$.
Obviously, we have $E_{t}=3N_{t}-6$. Moreover, according to the
connection rule, arbitrary two edges in the ESNs never cross each
other. Thus, the class of networks under consideration are maximal
planar networks (or graphs)~\cite{We01}.

\section{Relevant characteristics of the Networks}

In the following we will study the topology properties of $EW(t)$,
in terms of degree distribution, clustering coefficient, and average
path length.

\subsection{Degree distribution}

When a new node $i$ is added to the network at step $t_{i}$
$(t_{i}\geq 1)$, it has a degree of $4$. Let $L_{\vartriangle}(i,t)$
be the expected number of active triangles at step $t$ that will
create new nodes connected to the node $i$ at step $t+1$. Then at
step $t_{i}$, $L_{\vartriangle}(i,t_{i})=1 $. From the iterative
generation process of the network, one can see that at any
subsequent step each two new neighbors of $i$ generate two new
active triangles involving $i$, and one of its existing active
triangles is deactivated simultaneously. We define $k_{i}(t)$ as the
degree of node $i$ at time $t$, then the relation between $k_{i}(t)$
and $L_{\vartriangle}(i,t)$ satisfies:
\begin{equation}\label{activeTriangle}
L_{\vartriangle}(i,t)=\frac{k_{i}(t)-2}{2}.
\end{equation}

\begin{figure}[htb]
\begin{center}
\includegraphics[scale=0.8]{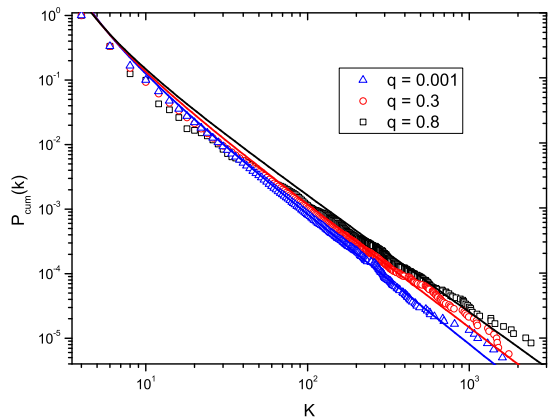}
\end{center}
\caption[kurzform]{\label{cumulateDegree} (Color online) The
cumulative degree distribution $P_{\rm cum}(k)$ at various $q$
values. The evolutionary steps of simulated network for $q=0.8$,
$q=0.3$, and $q=0.001$, are $t=12$, $t = 26$, and $t = 200000$,
respectively. The three lines are the theoretical results as
provided by Eq.~(\ref{gammak}). All data are from the average of ten
independent simulations.}
\end{figure}

Now we compute $L_{\vartriangle}(i,t)$. By construction,
$L_{\vartriangle}(i,t)=(1+q)L_{\vartriangle}(i,t-1)$. Considering
the initial condition $L_{\vartriangle}(i,t_{i})=1 $, we can derive
$L_{\vartriangle}(i,t)=(1+q)^{t-t_{i}}$. Then at time $t$, the
degree of vertex $i$ becomes
\begin{equation}\label{degree}
k_{i}(t)=2(1+q)^{t-t_{i}}+2.
\end{equation}
Since the degree of each node has been obtained explicitly as in
Eq.~(\ref{degree}), we can get the degree distribution via its
cumulative distribution \cite{Ne03}, i.e., $P_{\rm
cum}(k)=\sum_{k'\geq k}N(k',t)/N_{t}\sim k^{1-\gamma}$, where
$N(k',t)$ denotes the number of nodes with degree $k'$. The detailed
analysis is given as follows. For a degree $k=2(1+q)^{t-m}+2$, there
are $L_{v}(m)=3q(1+2q)^{m-1}$ nodes with this exact degree, all of
which were born at step $m$. All nodes born at time $m$ or earlier
have this or a higher degree. So we have
\begin{equation}
\sum_{k'\geq k}
N(k',t)=4+\sum_{s=1}^{m}L_{v}(s)=\frac{3(1+2q)^{m}+3}{2}.
\end{equation}
Thus, the cumulative degree distribution is give by
\begin{equation}
P_{\rm cum}(k)=\frac{1}{N_{t}}\sum_{k'\geq
k}N(k',t)=\frac{3(1+2q)^{m}+3}{3(1+2q)^{t}+3}.
\end{equation}
Substituting for $m$ in this expression using $m=t-\frac{\ln
\frac{k-2}{2}}{\ln (1+q)}$ gives
\begin{equation}
P_{\rm cum}(k)=\frac{3(1+2q)^{t}\cdot
2^\frac{\ln(1+2q)}{\ln(1+q)}(k-2)^{-\frac{\ln(1+2q)}{\ln(1+q)}}+3}{3(1+2q)^{t}+3}.
\end{equation}
When $t$ is large enough, one can obtain
\begin{equation}\label{gammak}
P_{\rm cum}(k)\approx
2^\frac{\ln(1+2q)}{\ln(1+q)}(k-2)^{-\frac{\ln(1+2q)}{\ln(1+q)}}.
\end{equation}
So the degree distribution follows a power-law form $P(k) \sim
k^{-\gamma}$ with the exponent
$\gamma=1+\frac{\ln(1+2q)}{\ln(1+q)}$. Note that the degree exponent
$\gamma$ is a continuous function of $q$, and belongs to the
interval $(1+\frac{\ln3}{\ln2},3]$. As $q$ decreases from 1 to 0,
$\gamma$ increases from $1+\frac{\ln3}{\ln2}$ to $3$. In the two
limitations, i.e., $q=1$ and $q\rightarrow0$ (but $q\neq 0$), the
evolutionary Sierpinski networks reduce to the deterministic
Sierpinski network~\cite{ZhZhChGu07} and its stochastic
variant~\cite{ZhZhSuZoGu08}, respectively. Figure
\ref{cumulateDegree} shows, on a logarithmic scale, the scaling
behavior of the cumulative degree distribution $P_{\rm cum}(k)$ for
different values of $q$. Numerical simulation agrees well with the
analytical result.

\subsection{ Clustering coefficient}

In a network, the clustering coefficient~\cite{WaSt98} $C_{i}$ of
node $i$ is defined as the ratio between the number of edges $e_{i}$
that actually exist among the $k_{i}$ neighbors of node $i$ and its
maximum possible value $k_{i}(k_{i}-1)/2$, i.e.,
$C_{i}=2e_{i}/[k_{i}(k_{i}-1)]$. The clustering coefficient of the
whole network is the average of $C'_{i}$s over all nodes in the
network.

For the ESNs, the analytical expression of clustering coefficient
$C(k)$ for a single node with degree $k$ can be derived exactly.
When a node is added into the network, its $k_{i}$ and $e_{i}$ are
both $4$. At each subsequent discrete time step, each of its active
triangles increases both $k_{i}$ and $e_{i}$ by $2$ and $3$,
respectively. Thus, $e_{i}=4+\frac{3}{2}(k_{i}-4)$ for all nodes at
all steps. So there is a one-to-one correspondence between the
clustering coefficient of a node and its degree. For a node of
degree $k$, we have
\begin{equation}\label{ACC}
C(k)=\frac{2\left[4+\frac{3}{2}(k-4)\right]}{k(k-1)}=\frac{4}{k}-\frac{1}{k-1},
\end{equation}
which is inversely proportional to $k$ in the limit of large $k$.
The scaling of $C(k)\sim k^{-1}$ has been empirically observed in
many real-life networks~\cite{RaBa03}.

\begin{figure}
\begin{center}
\includegraphics[width=0.6\textwidth]{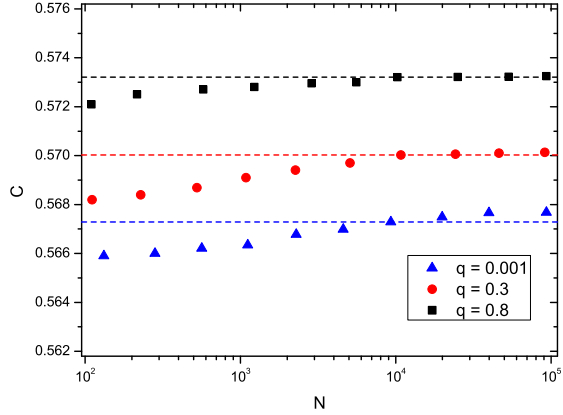} \
\end{center}
\caption[kurzform]{\label{cluster01} The clustering coefficient of
the whole network as a function of the size of the network for
various $q$. Results are averaged over ten network realizations for
each datum.}
\end{figure}

\begin{figure}
\begin{center}
\includegraphics[width=0.6\textwidth]{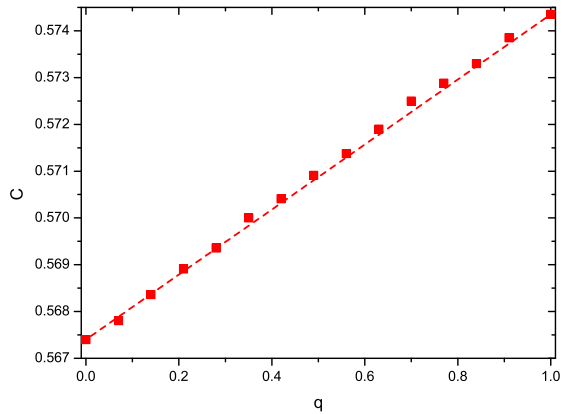}
\end{center}
\caption[kurzform]{\label{cluster02} The dependence of the average
clustering coefficient on parameter $q$.}
\end{figure}

After $t$ generation evolutions, the average clustering coefficient
$\bar{C}_{t}$ of network $EW(t)$ is given by
\begin{equation}\label{ACC1}
\bar{C}_{t}=\frac{1}{N_t}\sum_{r=0}^{t}\left[\left(\frac{4}{K_{r}}-\frac{1}{K_{r}-1}\right)L_{v}(r)\right],
\end{equation}
where the sum runs over all the nodes and $K_{r}$ is the degree of
those nodes created at step $r$, which is given by
Eq.~(\ref{degree}). In the infinite network order limit
($N_{t}\rightarrow \infty$), Eq.~(\ref{ACC1}) converges to a nonzero
value $C$, see Fig.~\ref{cluster01}. Moreover, it can be easily
proved that both $\bar{C}_{t}$ and $C$ increase with $q$. Exactly
analytical computation shows: when $q$ increases from $0$ to $1$,
$C$ grows from $0.5674$~\cite{ZhZhSuZoGu08} to
$0.5745$~\cite{ZhZhChGu07}. Therefore, the evolutionary networks are
highly clustered. Figure \ref{cluster02} shows the average
clustering coefficient of the network as a function of $q$, which is
in accordance with our above conclusions. From
Figs.~\ref{cumulateDegree} and \ref{cluster02}, one can see that
both degree exponent $\gamma$ and clustering coefficient $C$ depend
on the parameter $q$. The mechanism resulting in this relation
deserves further study. The fact that a biased choice of the active
triangles at each iteration may be a possible explanation, see
Ref.~\cite{CoRobA05}.

\subsection{Average path length}

From above discussions, one knows that the existing model shows both
the scale-free nature and the high clustering at the same time. In
fact, our model also possesses small-world property. Next, we will
show that our networks have at most a logarithmic average path
length (APL) with the number of nodes, where APL means the minimum
number of edges connecting a pair of nodes, averaged over all
couples of nodes.

Using a mean-field approach similar to that presented in
Refs.~\cite{ZhRoCo06,WaDaXu07}, one can predict the APL of our
networks analytically. By construction, at each time step, the
number of newly-created nodes is different. In order to distinguish
different nodes, we construct a node sequence in the following way:
when $\Delta N$ nodes are added at a given time step, we label them
as $M+1, M+2,\ldots, M+\Delta N$, where $M$ is the total number of
the pre-existing nodes. Eventually, every node is labeled by a
unique integer, and the total number of nodes is
$N_{t}=\frac{3(1+2q)^{t}+3}{2}$ at time $t$. We denote $L(N)$ as the
APL of the ESNs with order $N$. It follows that
$L(N)=\frac{2D(N)}{N(N-1)}$, where $D(N)=\sum_{1\leq i \leq j \leq
N}d_{i,j}$ is the total distance, and where $d_{i,j}$ is the
smallest distance between node $i$ and node $j$. Note that the
distances between existing node pairs are not affected by the
addition of new nodes. As in the analysis
of~\cite{ZhRoCo06,WaDaXu07}, we can easily derive that $D(N)\sim
N^{2}$ in the infinite limit of $N$. Then, $L(N)\sim \ln N$. Thus,
there is a slow growth of the APL with the network order $N$. This
logarithmic scaling of $L(N)$ with network order $N$, together with
the large clustering coefficient obtained in the preceding
subsection, shows that the considered graphs have a small-world
effect.

\begin{figure}[htb]
\begin{center}
\includegraphics[scale=0.8]{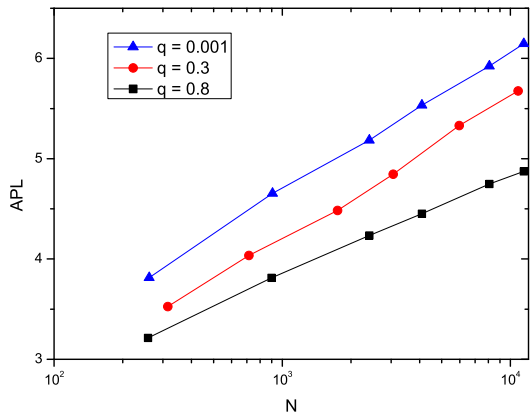}
\end{center}
\caption[kurzform]{\label{apl} Semilogarithmic graph of the APL
versus the network order $N$.}
\end{figure}

In particular, in the case of $q=1$, we can exactly compute the
average path length. A previously reported analytical result has
shown that the APL for this special case grows logarithmically with
the order of the network~\cite{ZhZhChGu07}. In Fig.~\ref{apl}, we
report the simulation results on the APL of ESNs for different $q$.
From Fig.~\ref{apl}, one can see that APL decreases with increasing
$q$. For all $q$, APL increases logarithmically with network order.

\section{Prisoner's Dilemma game on two limiting cases for ESN}

The ultimate goal of study of network structure is to study and
understand the workings of systems built upon those
networks~\cite{Ne03,BoLaMoChHw06}. Recently, some researchers have
focused on the analysis of functional or dynamical aspects of
processes occurring on networks. One particular issue attracting
much attention is using evolutionary game theory to analyze the
evolution of cooperation on different types of
networks~\cite{SzFa07}. Cooperation is ubiquitous in the real-life
systems, ranging from biological systems to economic and social
systems~\cite{Du97}. Understanding the emergence and survival of
cooperative behavior in these systems has become a fundamental and
central issue.

After studying the relevant characteristics of network structure,
which is described in the previous section, we will study the
evolutionary game behavior on the networks, with focus on the game
of prisoner's dilemma (PD). In the simple, one-shot PD game, both
receive $R$ under mutual cooperation and $P$ under mutual defection,
while a defector exploiting a cooperator gets amount $T$ and the
exploited cooperator receives $S$, such that $T>R>P>S$. As a result,
it is better to defect regardless of the opponent's decision, which
in turns makes cooperators unable to resist invasion by defectors,
and the defection is the only evolutionary stable strategy in fully
mixed populations.

We now investigate the evolutionary PD game on our networks to
reveal the influences of topological properties on cooperation
behavior. Here we only study two limiting cases: $q = 1$ and $q
\rightarrow 0$ (but $q \neq 0$). As usual in the
studies~\cite{NoMa92,SaPa05}, we choose the payoffs as $R=1$,
$P=S=0$, and $T=b>1$, and implement the finite population analogue
of replicator dynamics. During each generation, each individual $i$
plays the single given game with all its neighbors, and their
accumulated payoff being stored in $P_{i}$. After each round of the
game, the individual $i$ is allowed to update its strategy by
selecting at random a neighbor among all its neighbors, $j$, and
comparing their respective payoffs $P_{i}$ and $P_{j}$. If
$P_{i}>P_{j}$ , the individual $i$ will keep the same strategy for
the next generation. On the contrary, the individual $i$ will adopt
the strategy of its neighbor $j$ with a probability dependent on the
payoff difference ($P_{j}-P_{i}$) as $\prod_{i\rightarrow j}=(P_{j}-
P_{i})/\max\{k_{i},k_{j}\}b$. All individuals update its strategies
synchronously during the evolution process.

\begin{figure}[htb]
\begin{center}
\includegraphics[scale=0.8]{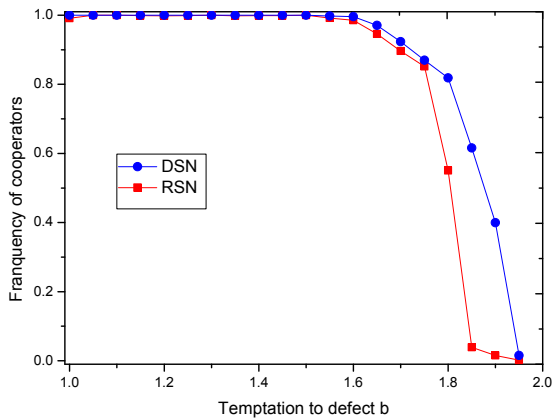}
\end{center}
\caption[kurzform]{\label{game} Frequency of cooperators as a
function of the temptation to defect $b$.}
\end{figure}

In Fig.~\ref{game}, we report the simulated results (i.e. the
dependence of the equilibrium frequency of cooperators on the
temptation to defect $b$) for both DSN and RSN. Simulations are
performed for both networks with order $9843$. Each data point is
obtained by averaging over 100 simulations for each of ten different
network realizations. From Fig.~\ref{game}, one can see that for $b<
1.6$ the cooperators in the deterministic Sierpinski network is
dominant over defectors. And similar phenomenon is also observed for
its corresponding random version. Thus, both of the network
structures are in favor of cooperation upon defection for a wide
range of $b$. Figure~\ref{game} also shows that in both networks the
frequency of cooperators makes a steady decrease when the temptation
changes from $1.6$ to $1.75$, and then drops dramatically when the
temptation increases to $2$. On the other hand, for large $b$ (such
as $b>1.6$), the equilibrium frequency of DSN is higher than that in
RSN, and this phenomenon is more obvious when the temptation is
larger and goes up to $2$.

The observed phenomena in Fig.~\ref{game} can be explained according
to the underlying network structures. Since both DSN and RSN are
scale-free networks, this heterogeneous network architecture makes
cooperation become the dominating trait over a wide range of
temptation to defect $b$~\cite{SaPa05,GoCaFlMo07}. Although both
networks have scale-free property, DSN is more heterogeneous than
RSN since the former has a smaller exponent of power-law degree
distribution than the latter; at the same time, the average
clustering coefficient of DSN is larger than that of RSN. These two
different characteristics between DSN and RSN can account for the
dissimilar cooperation behavior in both
networks~\cite{SaPa05,SaJeVi08}: A higher value of average
clustering coefficient, together with a smaller exponent of
power-law degree distribution, produces an overall improvement of
cooperation in DSN, even for a very large temptation to defect,
which is compared to that in RSN.


\section{Conclusions}

In summary, on the basis of Sierpinski gasket, we have proposed and
studied one kind of evolving network: evolutionary Sierpinski
networks (ESNs). According to the network construction process we
have presented an algorithm to generate the networks, based on which
we have obtained the analytical and numerical results for degree
distribution, clustering coefficient, as well as average path
length, which agree well with a large amount of real observations.
The degree exponent can be adjusted continuously between
$1+\frac{\ln3}{\ln2}$ and 3, and the clustering coefficient is very
large. Moreover, we have studied the evolutionary PD game on two
limiting cases of the ESN.

It should be stressed that the network representation introduced
here is convenient for studying the complexity of some real systems
and may have wider applicability. For instance, a similar recipe has
been recently adopted for investigating the navigational complexity
of cities~\cite{RoTrMiSn05}; on the other hand, it is frequently
used in RNA folding research~\cite{HsSt99,KaSt04}; moreover, earlier
links associating this network representation with polymers have
proven useful to the study of polymer
physics~\cite{VeSuKaLe99,KaSt05}. Thus, our study provides a
paradigm of representation for the complexity of many real-life
systems, making it possible to study the complexity of these systems
within the framework of network theory.

Because of its three important properties: power-law degree
distribution, small intervertex separation, and large clustering
coefficient, the proposed networks possess good structural features
in accordance with a variety of real-life networks. Additionally,
our networks are maximal planar graphs, which may be helpful for
designing printed circuits~\cite{We01}. Finally, it should be
mentioned that although our model can reproduce a few topological
characteristics of real-life systems, it remains unknown whether the
model can capture a true underlying mechanism responsible for those
properties observed in real networks. This belongs to the issue of
model evaluation, which is beyond the scope of the present paper but
deserves further study in future~\cite{CaDe08}.

\subsection*{Acknowledgment}

We thank Yichao Zhang for preparing this manuscript. This research
was supported by the National Basic Research Program of China under
grant No. 2007CB310806, the National Natural Science Foundation of
China under Grant Nos. 60704044, 60873040 and 60873070, Shanghai
Leading Academic Discipline Project No. B114, and the Program for
New Century Excellent Talents in University of China (NCET-06-0376).



\begin{thebibliography}{10}

\bibitem{AlBa02} R. Albert and A.-L. Barab\'asi,
       Rev. Mod. Phys. {\bf 74}, 47 (2002).

\bibitem{DoMe02} S.N. Dorogovtsev and J.F.F. Mendes,
Adv. Phys. {\bf 51}, 1079 (2002).

\bibitem{Ne03} M.E.J. Newman,
SIAM Rev. {\bf 45}, 167 (2003).

\bibitem{BoLaMoChHw06}
S. Boccaletti, V. Latora, Y. Moreno, M. Chavez and D.-U. Hwanga,
Phys. Rep. {\bf 424}, 175 (2006).

\bibitem{BaAl99} A.-L. Barab\'asi and R. Albert,
       Science {\bf 286}, 509 (1999).

\bibitem{WaSt98} D.J. Watts and H. Strogatz,
        Nature (London) {\bf 393}, 440 (1998).


\bibitem{ZhZhChGu07}
Z. Z. Zhang, S. G. Zhou, T. Zou, L. C. Chen, and J. H. Guan, Eur.
Phys. J. B {\bf 60}, 259 (2007).

\bibitem{ZhZhSuZoGu08}
Z. Z. Zhang, S. G. Zhou, Z. Su, T. Zou, and J. H. Guan, Eur. Phys.
J. B {\bf 65}, 141 (2008).


\bibitem{Si15}
 W. Sierpinski, Comptes Rendus (Paris) {\bf 160}, 302
(1915).

\bibitem{Re94}
C. A. Reiter, Comput. $\&$ Graphics {\bf 18}, 885 (1994).

\bibitem{We01} D.B. West,
    {\em Introduction to Graph Theory}
    (Prentice-Hall,  Upper Saddle River, NJ, 2001).

\bibitem{RaBa03}
E. Ravasz, A.-L. Barab\'asi, Phys. Rev. E 67, 026112 (2003).

\bibitem{CoRobA05}
F. Comellas, H. D. Rozenfeld, D. ben-Avraham,
Phys. Rev. E {\bf 72}, 046142 (2005).

\bibitem{ZhRoCo06}
Z. Z. Zhang, L. L. Rong, and F. Comellas, Physica A {\bf 364}, 610
(2006).

\bibitem{WaDaXu07}
L. Wang, H. P. Dai, and Y. X. Sun, J. Phys. A: Math. Thero. {\bf
40}, 13279 (2007).


\bibitem{SzFa07}
G. Szab\'o and G. F\'ath,
Phy. Rep. {\bf 446}, 97 (2007).

\bibitem{Du97}
L. A. Dugatkin, \emph{Cooperation Among Animals: An Evolutionary
Perspective} (Oxford University Press, Oxford, 1997).

\bibitem{NoMa92}
M. Nowak and R. May, Nature (London) {\bf 359}, 826 (1992).

\bibitem{SaPa05}
 F. C. Santos and J. M. Pacheco, Phys. Rev. Lett. {\bf 95},
098104 (2005).

\bibitem{GoCaFlMo07}
J. Gomez-Gardenes, M. Campillo, L. M. Floria, and Y. Moreno, Phys.
Rev. Lett. {\bf 98}, 108103 (2007).

\bibitem{SaJeVi08}
S. Assenza, J. G\'omez-Garde\~nes, and V. Latora, Phys. Rev. E {\bf
78}, 017107 (2008).

\bibitem{RoTrMiSn05}
M. Rosvall, A. Trusina, P. Minnhagen, and K. Sneppen, Phys. Rev.
Lett. {\bf 94}, 028701 (2005)

\bibitem{HsSt99}
C. Haslinger and P.F. Stadler,
Bull. Math. Biol. {\bf 61}, 437 (1999).

\bibitem{KaSt04}A. Kabak\c c\i o\u glu and  A.L. Stella,
Phys. Rev. E 70, 011802 (2004).

\bibitem{VeSuKaLe99} M. Vendruscolo, B. Subramanian, I. Kanter, E. Domany, and J. Lebowitz,
Phys. Rev. E {\bf 59}, 977 (1999).

\bibitem{KaSt05}
A. Kabak\c c\i o\u glu and  A.L. Stella,
Phys. Rev. E 72, 055102(R) (2005).


\bibitem{CaDe08}
A. Cami, N. Deo, Networks, {\bf 51}, 211 (2008).

\end{thebibliography}
\end{document}